\documentclass[conference,a4paper]{IEEEtran}
\IEEEoverridecommandlockouts

\usepackage{cite}
\usepackage{amsmath,amssymb,amsfonts}
\usepackage{algorithmic}
\usepackage{graphicx}
\usepackage{textcomp}
\usepackage{xcolor}
\usepackage{fancyhdr}
\usepackage{url}
\usepackage{multirow}
\usepackage{todonotes}
\usepackage{svg}
\usepackage{subcaption}
\newcommand{\oz}[1]{\textcolor{black}{#1}}
\newcommand{\tf}[1]{\textcolor{black}{#1}}

\usepackage[nolist]{acronym}
\newif\ifacronymlist
\acronymlisttrue

\makeatletter
\@ifundefined{nolist}{}{\acronymlistfalse}
\makeatother

\ifacronymlist
  \begin{acronym}[DSTTDSGRC]
\acro{AE}{Autoencoder}
\acro{AOA}{Angle Of Arrival}

\acro{DBSCAN}{Density-Based Spatial Clustering of Applications with Noise}
\acro{DL}{Deep Learning}

\acro{GRU}{Gated Recurrent Unit}

\acro{LSTM}{Long Short Term Memory}
\acro{MPCs}{MultiPath Components}
\acro{PDP}[PDP]{Power Delay Profile}

\acro{RNN}{Recurrent Neural Networks}
\acro{RL}{Reinforcement learning}
\acro{SVD}[SVD]{Singular Value Decomposition}
\acro{SVDD}{Support Vector Data Descriptor}
\acro{SVM}{Support Vector Machine}
\acro{TOA}{Time of Arrival}

\acro{USB}[USB]{Universal Serial Bus}
\acro{VAE}{Variational Auto-Encoder}
\acro{SNR}{Signal to Noise Ratio}
\end{acronym}
\fi

\def\BibTeX{{\rm B\kern-.05em{\sc i\kern-.025em b}\kern-.08em
    T\kern-.1667em\lower.7ex\hbox{E}\kern-.125emX}}

\fancypagestyle{firstpage}
{
    \fancyhf{}
    \fancyhead[C]{The paper has been presented at the 2025 35th International Conference Radioelektronika (RADIOELEKTRONIKA),\\ Hnanice, Czech Republic, May 12-14, 2025, \url{https://doi.org/10.1109/RADIOELEKTRONIKA65656.2025.11008404}}
    \fancyfoot[C]{This research was funded in part by the National Science Center (NCN), Poland, grant no. 2021/43/I/ST7/03294 (MubaMilWave). For this purpose of Open Access, the author has applied a CC-BY public copyright license to any Author Accepted Manuscript (AAM) version arising from this submission.}
}

\begin{document}

\title{A Deep Learning Approach to Multipath Component Detection in Power Delay Profiles\\
}
\author{\IEEEauthorblockN{1\textsuperscript{st} Ondrej Zeleny}
\IEEEauthorblockA{\textit{Dept. of Radio Electronics} \\
\textit{Brno University of Technology}\\
Brno, Czech Republic \\
ondrej.zeleny@vut.cz}
\and
\IEEEauthorblockN{2\textsuperscript{nd} Radek Zavorka}
\IEEEauthorblockA{\textit{Dept. of Radio Electronics} \\
\textit{Brno University of Technology}\\
Brno, Czech Republic \\
}
\and
\IEEEauthorblockN{3\textsuperscript{rd} Ales Prokes}
\IEEEauthorblockA{\textit{Dept. of Radio Electronics} \\
\textit{Brno University of Technology}\\
Brno, Czech Republic \\
}
\and
\IEEEauthorblockN{4\textsuperscript{th} Tomas Fryza}
\IEEEauthorblockA{\textit{Dept. of Radio Electronics} \\
\textit{Brno University of Technology}\\
Brno, Czech Republic \\
}
\and
\IEEEauthorblockN{5\textsuperscript{th} Wojtuń Jarosław}
\IEEEauthorblockA{\textit{Institute of Communications Systems} \\
\textit{Military University of Technology}\\
Warsaw, Poland \\
}
\and
\IEEEauthorblockN{6\textsuperscript{th} Jan M. Kelner}
\IEEEauthorblockA{\textit{Institute of Communications Systems} \\
\textit{Military University of Technology}\\
Warsaw, Poland \\
}
\and
\IEEEauthorblockN{7\textsuperscript{th} Cezary Ziółkowski}
\IEEEauthorblockA{\textit{Institute of Communications Systems} \\
\textit{Military University of Technology}\\
Warsaw, Poland \\
\and
\IEEEauthorblockN{8\textsuperscript{th} Aniruddha Chandra}
\IEEEauthorblockA{\textit{National Institute of Technology} \\
Dargapur, India \\
}
}

}

\maketitle
\thispagestyle{firstpage}

\begin{abstract}
\ac{PDP} plays a crucial role in wireless communications, providing information on multipath propagation and signal strength variations over time. Accurate detection of peaks within \ac{PDP} is essential to identify dominant signal paths, which are critical for tasks such as channel estimation, localization, and interference management. Traditional approaches to \ac{PDP} analysis often struggle with noise, low resolution, and the inherent complexity of wireless environments.

In this paper, we evaluate the application of traditional and modern deep learning neural networks to reconstruction-based anomaly detection to detect multipath components within the \ac{PDP}. To further refine detection and robustness, a framework is proposed that combines autoencoders and \ac{DBSCAN} clustering. To compare the performance of individual models, a relaxed F1 score strategy is defined. The experimental results show that the proposed framework with transformer-based autoencoder shows superior performance both in terms of reconstruction and anomaly detection. 

\end{abstract}


\section{Introduction}
Wireless communication systems rely on the transmission of signals through various propagation environments, where signals interact with obstacles, reflect \tf{off} surfaces, and scatter before reaching the receiver. This phenomenon, known as multipath propagation, significantly affects signal quality, leading to fading, interference, and time dispersion. Understanding and accurately characterizing multipath effects is crucial for optimizing wireless system performance, improving localization accuracy, and overall network efficiency.

The key characteristic for the analysis of multipath propagation is the \ac{PDP} \tf{(Power Delay Profile)}, which represents the power of the received signal as a function of time delay. The peak in the PDP corresponds to the dominant signal paths, providing valuable information about the structure of the propagation environment. Detecting these peaks is essential for multiple applications in wireless communications, including:
\begin{itemize}
    \item Channel Estimation: Identifying key \ac{MPCs} enables better modeling of the wireless channel, improving equalization, and signal recovery.
    \item 5G/6G Positioning: Accurate PDP analysis improves the estimation of \ac{TOA} and \ac{AOA}, crucial for high-precision localization.
    \item Adaptive Modulation and Coding: Reliable detection of multipath components allows dynamic adjustment of transmission parameters, optimizing data rates, and minimizing errors.
\end{itemize}

Traditional approaches to \ac{PDP} peak detection, such as thresholding-based methods, matched filtering, and maximum likelihood estimation, often struggle in complex wireless environments. These methods are sensitive to noise, require extensive tuning, and may fail to distinguish weak \ac{MPCs} from background interference. Furthermore, as wireless networks transition to higher frequency bands (e.g., mmWave, THz), propagation effects become more pronounced, necessitating more robust analytical techniques~\cite{SHANG20131317, Wang691, 10.1145/3310194}.

Recent advances in \ac{DL} have introduced powerful tools for modeling complex patterns in signal processing tasks. Modern neural architectures, such as Transformers and attention mechanisms, provide an opportunity to enhance \ac{PDP} analysis by capturing long-range dependencies and learning contextual relationships between \ac{MPCs}. Unlike conventional algorithms, these models can adapt to diverse environments, generalize across different channel conditions, and improve peak detection accuracy~\cite{Pang_2021}.

\oz{
This paper aims to explore the applications of deep learning for detecting \ac{MPCs} in \ac{PDP}. To achieve this, reconstruction-based deep learning models, including autoencoders and transformers, are proposed to enhance peak detection, even at low \ac{SNR}.
The paper is structured as follows. In Section~\ref{Theory}, the principle of different autoencoders is outlined. The following \tf{S}ection~\ref{State of The Art}, presents studies that address similar problems. In Section~\ref{Proposal}, a~framework for peak detection in \ac{PDP} based on deep learning is proposed. The experimental results of the proposed framework are presented in Section~\ref{Experimental Results}. Lastly, Section~\ref{Conclusion} summarizes the experimental results and outlines the future work.
}

\section{Theory}\label{Theory}

\begin{figure}
    \begin{center}
        \includegraphics[width=\columnwidth]{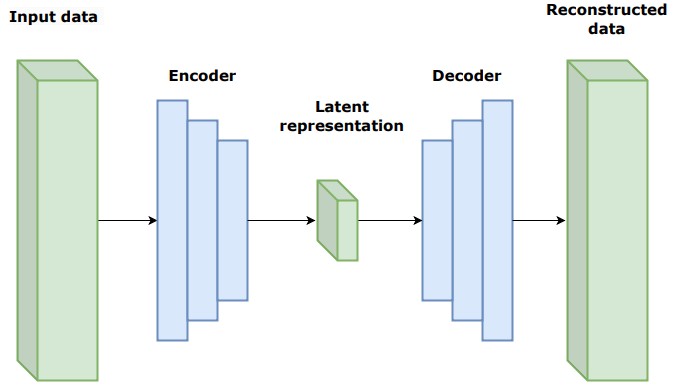}
        \caption{\oz{Autoencoder.}}
        \label{autoencoder}
    \end{center}
\end{figure}

Deep learning approaches, particularly those based on reconstruction-based learning, have gained popularity due to their ability to model complex temporal dependencies and capture deviations from normal behavior. This section explores key deep learning models that rely on reconstruction, including \ac{AE}, \ac{GRU}, \ac{LSTM} networks, and transformer-based models. Traditional \ac{AE} are a type of fully connected artificial neural network designed for unsupervised learning. Their primary function is to encode the input data into a compressed representation and then reconstruct it as accurately as possible. They learn by minimizing the reconstruction error, ensuring that normal data points are well reconstructed while anomalies, which typically deviate from the learned patterns, result in higher reconstruction errors~\cite{Pang_2021}. The architecture of the autoencoder is illustrated in Figure~\ref{autoencoder}.

CNN-based autoencoders extend the traditional autoencoder architecture by incorporating convolutional layers~\cite{DBLP:journals/corr/OSheaN15} instead of fully connected layers. These models are particularly effective in capturing spatial hierarchies and local patterns in data, making them well-suited for applications involving structured inputs such as images, time series, and spatial-temporal data. In the context of time-series anomaly detection, \tf{the} CNN-based autoencoders can be used to learn representative temporal patterns while maintaining local dependencies. By applying one-dimensional convolutions, these models effectively capture short- and long-term dependencies without requiring recurrent connections, making them computationally efficient compared to RNN-based approaches. The reconstruction error serves as a measure to detect anomalies, where deviations from expected patterns indicate potential outliers~\cite{THILL2021107751}.

\ac{RNN}-based autoencoders are widely used to model sequential data, as they incorporate temporal dependencies through recurrent connections. Unlike feedforward networks, \ac{RNN}'s maintain hidden states that capture information from previous time steps, making them effective for time-series modeling and anomaly detection. However, standard RNNs suffer from vanishing-gradient problems, which limits their ability to model long-range dependencies. To address this issue, advanced variants such as \ac{LSTM} networks and \ac{GRU} are typically used. In reconstruction-based anomaly detection, \ac{LSTM}-based autoencoders encode sequential data into a compact latent representation while preserving temporal dependencies. The decoder reconstructs the sequence and anomalies are identified on the basis of high reconstruction errors. LSTMs are particularly effective in handling time series with complex temporal patterns, making them a preferred choice for industrial monitoring, financial fraud detection, and physiological signal analysis~\cite{wei2022lstmautoencoderbasedanomalydetection, Abdelli_2021}.

Transformer-based models, such as the Anomaly Transformer, have recently gained attention for anomaly detection due to their ability to model long-range dependencies efficiently. Unlike RNNs and LSTMs, Transformers rely on self-attention mechanisms, allowing them to capture global contextual information without the constraints of sequential processing. Transformers can be used in autoencoder frameworks, where an encoder learns meaningful representations from the input, and a decoder attempts to reconstruct it. Self-attention mechanisms enable Transformers to focus on the most relevant parts of the sequence, improving their ability to detect subtle anomalies. Moreover, transformers can take advantage of association discrepancy learning, a technique that differentiates between normal and anomalous patterns by measuring attention divergence in different time steps~\cite{10164104, MENTZELOPOULOS2024118639}. 

\section{State of The Art}\label{State of The Art}
A study by Zhang et al.~\tf{\cite{9665207}} (2021) introduced a machine learning approach to cluster \ac{MPCs} in high-speed railway communication environments. The methodology involves extracting features from MPCs and applying clustering algorithms to identify distinct propagation paths. This approach improves the understanding of channel characteristics, leading to improved design and optimization of communication systems in dynamic scenarios. 

In 2023, researchers proposed a hybrid receiver architecture combining artificial intelligence/machine learning (AI/ML) models with traditional peak detection methods for the 5G New Radio Physical Random Access Channel (PRACH)~\tf{\cite{singh2024machinelearningbasedhybrid}}. The AI/ML model processes combined \ac{PDP} from multiple antennas to detect user presence, followed by conventional peak detection for timing advance estimation. This hybrid approach addresses challenges in high-fading and low signal-to-noise ratio conditions, reducing false peaks and missed detection.

In 2020, Munin et al.~\tf{\cite{munin:hal-02359943}} explored the application of convolutional neural networks (CNNs) to detect multipath effects in Global Navigation Satellite System (GNSS) signals. By mapping the output signals of the correlator to two-dimensional images, CNN automatically extracts the relevant characteristics to identify multipath contamination. This technique improves positioning accuracy by effectively distinguishing between line-of-sight and multipath-affected signals.

Later, in 2022, the study by Pan et al.~\tf{\cite{Pan2023-ns}} proposed a machine learning-based method for multipath mitigation by formulating multipath modeling as a regression task. The approach utilizes spatial domain information to predict and compensate for multipath-induced errors, improving the reliability of the wireless communication system.

In 2024, the authors of the paper "Deep Learning for Multipath Detection in Ultra-Low Altitude Targets"~\tf{\cite{rs16244773}} presented deep learning techniques to detect ultra-low altitude targets affected by multipath propagation. By defining both the target and its multipath reflections as generalized targets in Range-Doppler maps, the study employed deep learning models to learn features of these generalized targets, enhancing detection capabilities in complex environments.

\section{\tf{Proposed Framework}}\label{Proposal}
\begin{figure}[h]
    \begin{center}
        \includegraphics[width=\columnwidth]{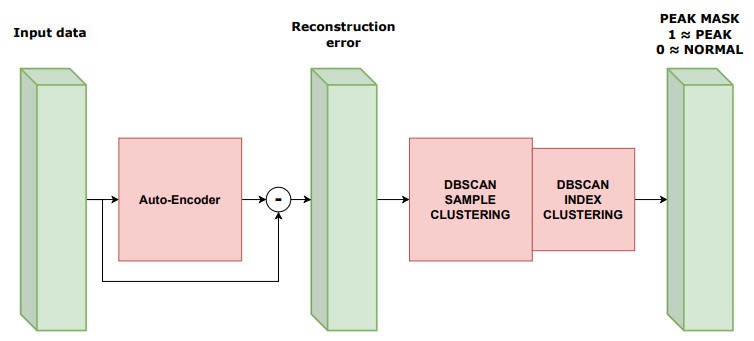}
        \caption{\oz{Proposed \tf{peak detection f}ramework.}}
        \label{Blocks}
    \end{center}
\end{figure}

\oz{
The proposed \tf{peak} detection method consists of two stages. First, an autoencoder is trained to reconstruct the \ac{PDP} signal, learning its general pattern while struggling to reproduce sudden spikes, that correspond to \ac{MPCs}. This results in higher reconstruction errors at these points. Once trained, the autoencoder is frozen and its reconstruction errors are passed to the second stage, where the \acl{DBSCAN} algorithm is applied.}
\oz{
Before clustering, negative reconstruction errors are filtered out to prevent unwanted noise. The first DBSCAN pass then groups high-error areas, identifying potential peaks. However, since multiple points may belong to the same peak, a second DBSCAN clustering step is applied to refine detections, ensuring that only the most significant peak per anomaly is retained. Finally, the detected peak indices in each cluster are matched with the original data, and only the peaks with the highest power are retained. Figure~\ref{Blocks} shows a simplified diagram of the entire detection method.
}

\section{Experimental Results}\label{Experimental Results}
\begin{figure}
    \begin{center}
        \includegraphics[width=\columnwidth]{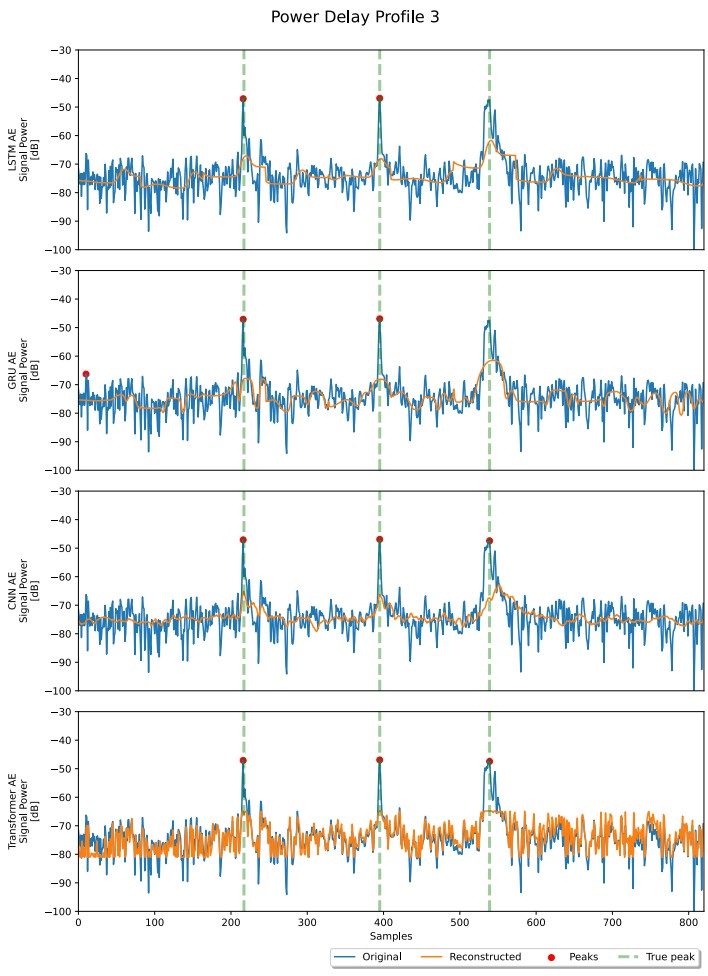}
        \caption{\oz{Peak Detection Results for the \ac{PDP} Sequence 3 from the Dataset.}}
        \label{sample0}
    \end{center}
\end{figure}

\begin{figure}
    \begin{center}
        \includegraphics[width=\columnwidth]{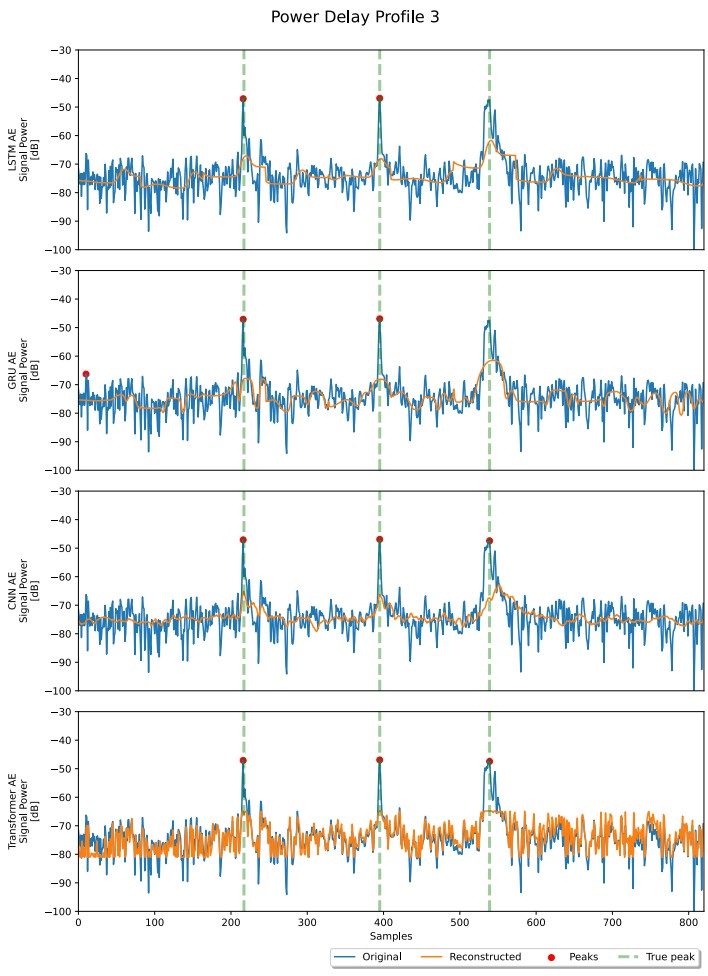}
        \caption{\oz{Peak Detection Results for the \ac{PDP} Sequence 26 from the Dataset.}}
        \label{sample1}
    \end{center}
\end{figure}

\begin{table}[!h]
\caption{Autoencoder Configuration.}
\begin{center}
\begin{tabular}{|l|p{0.28in}|p{0.6in}|p{0.4in}|p{0.3in}|p{0.3in}|}
\hline
Autoencoder            & Layers & Embedding / Channels      & Attention Heads & Kernel size \\ \hline
CNN         & 4      & 64, 128, 256, 512 & -               & 14          \\ \hline
LSTM        & 4      & 64, 32, 16, 8     & -               & -           \\ \hline
GRU         & 4      & 64, 32, 16, 8     & -               & -           \\ \hline
TRANSFORMER & 1      & 8                 & 1               & -           \\ \hline
\end{tabular}
\label{tab:configuration}
\end{center}
\end{table}
\oz{
The data used in the experiments come from previous research~\cite{osti_10165863}, which focused on measuring and analyzing signal propagation in a 60 GHz communication channel between a moving vehicle and a fixed receiver. The dataset consists primarily of \ac{PDP} measurements, capturing how signal strength varies over time due to reflections from roads, buildings, trees, and other objects. Two scenarios with a total of 80 \ac{PDP} measurements were selected for this study.
}
\oz{
Each sample was labeled with binary values indicating the presence of the most dominant and recognizable multipath components. The dataset was divided into training and testing sets, with alternating samples (e.g., samples with odd and even indices) assigned to each set. This method ensures that both sets contain samples evenly distributed over time. Given the relatively small dataset size, extensive data augmentation was applied after this split, including rolling shifts and Gaussian noise addition, to increase sample diversity. These techniques help improve model generalization, reduce overfitting, and preserve key multipath characteristics.
}
\oz{
We evaluated four deep learning \ac{AE} models: convolutional autoencoders (CNN-AE), LSTM autoencoders (LSTM-AE), GRU autoencoders (GRU-AE), and transformer-based autoencoders. The implementation was done using the torch library and all the necessary parameters are listed in Table~\ref{tab:configuration}. The optimization of individual models was done using ADAM optimizer, learning rate $0.001$ and trained for 50 epochs with an early stop if the models started to overfit. The positional encoding used for the transformer is sine-cosine, the same as in the original paper~\cite{DBLP:journals/corr/VaswaniSPUJGKP17}. The training process was the same for all models except for the LSTM- and GRU-based autoencoders, which struggle with long sequences due to vanishing gradients. To mitigate this, we followed~\cite{10365265} and split the \ac{PDP} sequences into shorter subsequences of lengths $410$, $205$, $85$, and $41$ (corresponding to $1/2$, $1/4$, $1/10$, and $1/20$ of the original length) to assess the reconstruction capabilities of LSTM and GRU networks at different sequence lengths. The sequences of length $410$ and $205$ remained too long for effective learning. At length $41$, both models were able to learn but often reconstructed peaks of \ac{MPCs}, which was undesirable. The optimal balance was found at a sequence length of $85$, where the models were effectively learned without excessive reconstruction. Consequently, this sequence length was used for further optimization.
}
\begin{table}[htbp]
\caption{Results of Tested Autoencoders.}
\begin{center}
    \begin{tabular}{|l|l|l|l|l|}
        \hline
                  & CNN & GRU & LSTM & TRANSFORMER \\ \hline
        Recall    & \tf{0}.53 & \tf{0}.44 & \tf{0}.55  & 0.61        \\ \hline
        Precision & \tf{0}.42 & \tf{0}.46 & \tf{0}.42  & \tf{0}.73        \\ \hline
        F1        & \tf{0}.47 & \tf{0}.45 & \tf{0}.48  & \tf{0}.66       \\
        \hline
        \end{tabular} 
    \label{tab:metrics}
\end{center}
\end{table}
\oz{
Instead of relying solely on visual evaluation, we applied relaxed recall, precision, and F1-score metrics, incorporating a tolerance window of $N$ steps around each labeled anomaly. These metrics provide meaningful insight into model performance; however, higher scores do not always indicate better results. In particular, recall must be interpreted with caution, as signal labels may not fully capture all multipath components. High recall suggests that most multipath components were detected, but it can also lead to more false positives, reducing precision. In real applications, balancing recall and precision is crucial to avoid detecting irrelevant components while ensuring that all significant signal paths are identified. The summary of the final results are shown in Table~\ref{tab:metrics}. Example samples from the testing phase are shown in Figures~\ref{sample0} and~\ref{sample1}\tf{, respectively}, where we can see the original \ac{PDP} signal (blue), the reconstructed \ac{PDP} signal (orange), the detected \ac{MPCs} peaks (red dots) and the ground truths used for the metrics (green vertical lines).
}

\section{Conclusion}\label{Conclusion}
\oz{
In this paper, we proposed a deep learning-based approach for detecting multipath components in \ac{PDP}. The method utilizes autoencoders to reconstruct PDP signals, with reconstruction errors serving as indicators of peak locations. The integration of DBSCAN clustering enables the identification of dominant multipath components while filtering out noise and irrelevant peaks.
}
Figures~\ref{sample0} and~\ref{sample1} show that the reconstruction of \ac{PDP} can be done reasonably well with all the autoencoders, however, the transformer AE shows exceptional performance in reconstructing the noise in the \ac{PDP}, thus minimizing noise in the reconstruction error space.
\oz{
Experimental results also show that transformer-based autoencoders achieve the highest recall (61\%) and F1-score (66\%), while LSTM-based models provide a more balanced performance with a recall of 55\% and an F1-score of 48\%. CNN and GRU-based autoencoders perform slightly worse in terms of recall but demonstrate competitive precision. These results highlight the challenges of multipath component detection in complex wireless environments while demonstrating the potential of deep learning methods to improve detection compared to traditional heuristics. The relaxed F1-score evaluation framework, which accounts for slight deviations in peak positions, provides a fairer comparison of different architectures and helps mitigate the impact of annotation uncertainties.
}
\oz{
Although this approach has limitations in terms of recall, it offers a structured and data-driven way to analyze multipath components without relying on manually tuned heuristics. Additionally, the relatively small dataset required extensive augmentation to improve model generalization. Future work will aim to improve recall while maintaining precision by improving the model's ability to distinguish meaningful patterns from noise. Exploring alternative representations and learning strategies, such as \ac{RL}, could further improve performance. In addition, testing in more complex and diverse scenarios will provide a better assessment of robustness and generalization. In addition, incorporating domain knowledge, advanced \ac{RL} techniques, and real-world constraints may further refine accuracy and enhance real-world applicability.
}



\section*{Acknowledgment}

The research described in this paper was supported by the Internal Grant Agency of the Brno University of Technology under project no. FEKT-S-23-8191, Czech Science Foundation (GACR) project no. 23-04304L and by the National Science Centre, Poland under research project no. 2021/43/I/ST7/03294 acronym ‘MubaMilWave’.

\bibliographystyle{IEEEtran}
\bibliography{references}{}

\end{document}